\documentclass[12pt,preprint]{aastex}

\shorttitle{Planetesimal Formation in Inhomogeneous MRI Disk}
\shortauthors{Kato et al.}

\begin{document}

\title{Planetesimal Formation at the Boundary Between Steady
Super/Sub-Keplerian Flow Created by Inhomogeneous Growth of Magnetorotational Instability}

\author{M. T. Kato}
\affil{Department of Earth and Planetary Science, Tokyo Institute of Technology, Ookayama 2-1-12-I2-10, Meguro-ku, Tokyo}

\author{M. Fujimoto}
\affil{Institute of Space and Astronomical Science, Japan Aerospace Exploration Agency, Yoshinodai 3-1-1, Sagamihara, Kanagawa}

\and

\author{S. Ida}
\affil{Department of Earth and Planetary Science, Tokyo Institute of Technology, Ookayama 2-1-12-I2-10, Meguro-ku, Tokyo}
\email{ida@geo.titech.ac.jp}

\begin{abstract}

We have studied formation of planetesimals at a radial pressure bump in a protoplanetary disk
created by radially inhomogeneous magnetorotational instability (MRI), through
three-dimensional resistive MHD simulations including dust particles.
In our previous papers, we showed that
the inhomogeneous MRI developing in non-uniform structure of magnetic field or magnetic resistivity can 
transform the local gas flow in the disk to a quasi-steady state with local rigid rotation that is no more
unstable against the MRI.
Since the outer part of the rigid rotation is super-Keplerian flow, 
a quasi-static pressure bump is created and
dust concentration is expected there.
In this paper, we perform simulations of the same systems, 
adding dust particles that suffer gas drag
and modulate gas flow via the back-reaction of the gas drag (dust drag).
We use $\sim O(10^7)$ super-particles, each of which represents $\sim O(10^6)$--$O(10^7)$ 
dust particles with sizes of centimeter to meter.
We have found that the dust drag suppresses turbulent motion
to decrease the velocity dispersion of the dust particles
while it broadens the dust concentrated regions to limit peaky dust concentration, 
compared with the simulation without the dust drag.
We found that the positive effect for the gravitational instability 
(reduction in the velocity dispersion) is dominated over 
the negative one (suppression in particle concentration).
For meter-size particles with the friction time $\tau_f \simeq 1/\Omega$, 
where $\Omega$ is Keplerian frequency, the gravitational instability of the
dust particles that may lead to planetesimal formation is expected.
For such a situation, we further introduced the self-gravity of dust particles 
to the simulation to
demonstrate that several gravitationally bound clumps are actually formed.
Through analytical arguments, we found that the planetesimal formation 
from meter-sized dust particles can be possible 
at $\sim 5$AU, if
dust spatial density is a few times larger than that in the minimum mass solar nebula.
\end{abstract}


\defcitealias{kato09}{Paper~I}
\defcitealias{kato10}{Paper~II}

\section{Introduction}

Planets form from coalescence of planetesimals in a protoplanetary disk.
Planetesimals with more than kilometer sizes should form from
dust grains that are initially less than micrometer sizes.
However, so called "meter-size barrier" exists.
Because meter-size particles are marginally coupled with disk gas motion
and the disk gas rotates slightly slower than Keplerian motion due to
radially negative pressure gradient of the gas,
the particles suffer "headwind" and rapidly migrate toward the host star.
The infall timescale is only a few hundred years for meter-size particles 
\citep{weiden77, nakagawa81}, which is much shorter than 
the growth timescale of particles by mutual collisions.
It has not been clarified how the particles grow over meter-sizes before infalling to
the host star.

One way to bypass the meter-size barrier is to form clumps from dust particles through self-gravitational instability (GI), which occurs on orbital periods \citep{safronov69, gold73},
if the dust particles locally have a large enough spatial density. 
Once bodies of kilometer-size or more are formed, they no longer undergo rapid migration.
Original idea for dust concentration for onset of GI was vertical settling of
dust grains onto the disk midplane.
However, the dust settling induces
Kelvin-Helmholtz instability due to difference in rotation velocities 
between the dust-rich layer (Keplerian) and an overlaying dust-poor layer 
(sub-Keplerian), and it prevents the dust layer from becoming
dense enough for GI \citep[e.g.,][]{weiden80,cuzzi93,sekiya98,ishitsu03,barranco09}
unless initial dust to gas ratio in the disk is sufficiently high \citep[e.g.,][]{chiang08,lee10a,lee10b}. 

Other than the induced KH instability, global turbulence is likely to exist
in the disk.
While the turbulence generally scatters dust particles, 
it could concentrate dust particles in anti-cyclonic vortexes \citep[e.g.,][]{barge95, chavanis00, johan04, inaba06}.
For this mechanism to lead to GI,
relatively high initial surface density of dust and
very weak turbulence may be required.
In the case of strong turbulence,
dust particles have too high velocity dispersion for GI
and the high collision velocity between dust particles 
results in fragmentation rather than growth \citep{guttler09, zsom10}. 
\citet{johan07} performed local three-dimensional MHD simulation including dust particles 
and showed that weakly fluctuating pressure bumps are created by 
magnetorotational instability (MRI) 
and meter-size bodies are concentrated at the bumps.
In the relatively-weak turbulence with the viscosity $\alpha \sim 10^{-3}$, 
the dust particles could stay long enough and increase their density to cause GI.
They found that back-reaction of drag force from gas to the dust particles,
which we  hereafter call "dust drag force," modulates gas motion to follow the particles
in dust-accumulated regions and weaken the turbulence. 

Although vertical sedimentation of dust is inhibited by KH instability,
radial accumulation is possible.
For example, radial dependence of speed of dust migration due to gas drag can
enhance the dust to gas ratio to facilitate GI in inner disk regions 
\citep{youdin_shu02,youdin_chinag04}.
This radial migration induces "streaming instability" if dust drag is considered.
In local dust-accumulated areas (dust clumps),
the dust drag force modulates the gas flow closer to Keplerian rotation.
As a result, "head wind" to the clumps becomes weaker and their
radial migration due to the gas drag becomes slower, 
which leads to rapid growth of the clumps by capturing dust particles 
and smaller clumps that migrate faster from outer regions 
\citep[e.g.,][]{you05, you07, johan07:si, johan, bai10a, bai10b, bai10c}.
The suppression of local turbulence by the dust drag
decreases the velocity dispersion of the dust particles in the clumps, which is also favorable for the GI. 

A global radial pressure bump also leads to radial concentration of dust.
The inner boundary of "dead zone" is one of such locations.
The growth rate of MRI depends on the magnetic strength and the ionization degree of disk gas \citep[e.g.,][]{jin96, sano99}. 
In the region where the gas ionization degree is low enough or 
the vertical magnetic field is weak enough, the ohmic dissipation decays MRI there ("dead zone"). 
The disk gas is ionized by thermal ionization, the stellar X-rays \citep[e.g.][]{igea99}, the cosmic rays \citep[e.g.,][]{ume83} and the radionuclides \citep[e.g.][]{stepinski92}. 
\citet{gammie96} and \citet{sano00} showed that the dead zone 
exists in the disk and it is confined in the inner disk ($\la 10$AU) 
near the disk midplane.
Because the viscosity is lower in the dead zone and disk accretion flux is conserved 
between the dead and active zones, disk gas column density is enhanced in the dead zone.
The positive radial gradient of gas column density at the inner boundary
of the dead zone produces a pressure bump in which dust particles accumulate
\citep{dzy10}. 
However, the inner boundary may be located at $\lesssim 1$AU, so 
the planetesimal formation there may be unable to account for
formation of icy planets and cores of gas giants. 

The column density of tiny dust grains may be enhanced around a snow line due to slowdown of the dust radial migration speed by diffusion of sublimated vapor \citep{cuzzi08} or 
by down-sizing through sublimation of a icy mantle around a silicate core of dust particles \citep{saito11}. 
Since the ionization degree depends on the abundance of the tiny grains \citep{sano00}, the ionization degree would significantly decrease around the snow line to produce a local dead zone \citep{kretke07,ida08}.
The inner edge of the local dead zone is a favorable site for rapid dust growth in relatively
outer regions \citep{brauer08}.
\citet[][which are referred to as Paper I]{kato09} and \citet[][Paper II]{kato10}
pointed out that if the local dead zone induced by the snow line
is embedded in the global MRI active zone, the divided inner active zone is 
sandwiched by the inner global dead zone and the outer local dead zone
and it can be a stable barrier
for dust radial migration in which dust particles are accumulated.
Even if the snow line is not in the global active zone, 
near the outer boundary of dead zone where MRI is marginal, 
fluctuations of magnetic fields and/or ionization degree could make
radially nonuniform MRI structure that can be approximated by
an active zone radially sandwiched by dead zones.  

In Paper I and II, we have investigated evolution of gas flow of the
active zone radially sandwiched by dead zones, through
shearing-box magnetohydrodynamic simulations.
In Paper I,  performing two-dimensional simulations,
we found that the angular velocity profile of gas is modified by 
local MRI turbulence in radially non-uniform magnetic field.
The vigorous angular momentum and mass transport associated with
the MRI turbulence lead to a local rigid rotation in the originally active zone.
The MRI turbulence can decay to the viscosity level of 
$\alpha\sim 10^{-4}$ after the transformation to the
quasi-steady state with the local rigid rotation, because
there is no shear motion to create MRI while magnetic field remains.
In the outer part of the local rigid rotation, gas rotation is
super-Keplerian and a pressure bump is formed.
Note that this gas flow structure is stable.
If the rigid rotation is broken, the induced shear motion again
produces MRI and transports angular momentum and mass to
recover the rigid rotation as long as the strong enough magnetic field remains. 

\begin{figure}
\figurenum{1}
\epsscale{0.6}
\plotone{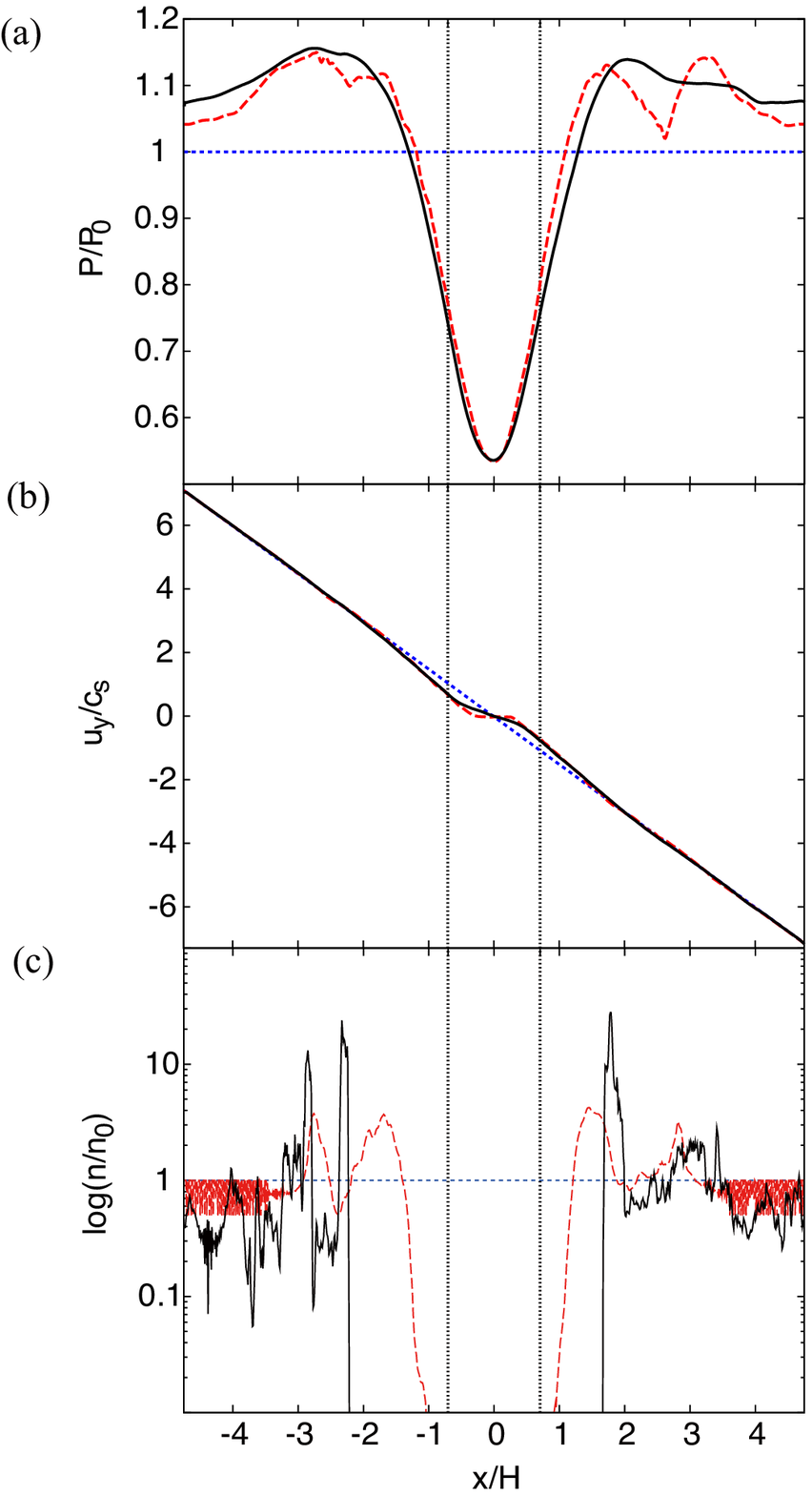}
\caption{Results of model-s40-t10-test described in Paper II. 
Time evolution of vertically averaged values of (a) pressure $P$, (b) gas angular velocity $u_{y}$ and (c) number density of particles $n$. 
$P$ and $n$ are normalized by the initial values
($P_0$ and $n_0$), and $u_y$ is normalized by sound speed $c_s$. 
The dotted, dashed and bold lines represent the snapshots at $t\Omega=0, 40$ and 70, respectively. 
The two vertical dotted-lines are the boundaries between the initially active (unstable) and dead (stable) regions. 
MRI is initially excited only in the region between the two dotted lines. }
\label{fig:paper2}
\end{figure}

In Paper II, we found the same local rigid rotation in the
three-dimensional (instead of two-dimensional) simulation as
shown in Figure~\ref{fig:paper2}b.
The calculations with test particles show that
the boundary region between sub- and super-Keplerian zones
acts as a strong and stable barrier for the dust migration and it
leads to dust particle concentration up to 10,000 times of the initial value
(Figure~\ref{fig:paper2}c),
which could eventually lead to planetesimal formation.

However, the dust drag force onto gas was neglected in Paper II,
although it would affect the dust concentration and velocity dispersion of dust particles
in the concentrated regions by altering the gas flow.
The drag lowers velocity dispersion of dust particles
to facilitate the GI, while it broadens the dust accumulated region
and suppresses peaky dust concentration that is rather negative for the GI. 
The latter effect is positive for the GI, while the former is negative. 
In this paper, we include the dust drag force to the simulations.
The equations and initial setup employed in our simulation are described 
in section~\ref{sec:equations}. 
In section~\ref{sec:drag}, we show the simulation results.
In section~\ref{sec:planetesimal}, we estimate the possibility of 
the planetesimal formation by analytical arguments.
We also demonstrate the planetesimal formation via the GI
by numerical simulation including the dust self-gravity. 
Section~\ref{sec:discussion} is devoted for conclusion and discussion.

\section{Equations and model}\label{sec:equations}

\subsection{Equations}

We consider a small region around the midplane which is rotating with the Keplerian frequency $\Omega$ at a distance $r$ from a central star to study local dust motion and magnetohydrodynamics. 
The coordinates that we use are $(x,y,z)$ where $x$ is the radial distance from $r$, $y$ is tangential distance, 
and $z$ is vertical distance from the disk midplane.

We include centimeter to meter-size dust particles as super-particles. 
Total number of the super-particles is $O(10^7)$ and 
each super-particle represents $O(10^6)$--$0(10^7)$ small dust particles.
The equation of motion of the $i$-th particle is given by
\begin{eqnarray}
\frac{{\rm d} \mathbf{v}_{i}}{{\rm d} t}&=&
-2\mathbf{\Omega}\times\mathbf{v}_{i}+3\Omega^2x_{i}\hat{\mathbf{x}}
-\frac{1}{\tau_{f}}\left(\mathbf{v}_{i}-\mathbf{u}\right)
-\nabla \Phi, \label{eq:particle}
\end{eqnarray}
where $\mathbf{u}$ is the gas velocity at the location of the $i$-th particle, which is interpolated using gas velocities at the neighbor grid points. 
The third term in the r.~h.~s. (right hand side) represents the specific gas drag force to the dust particle.
We adopt the Epstein drag force,  
\begin{eqnarray}
\tau_f=\rho_{p} a/(\rho_g c_s),
\end{eqnarray}
where $\rho_{p}$ and $a$ are the internal density and radius of dust particles, 
$\rho_{g}$ and $c_s$ are the spatial density and sound velocity of surrounding gas.  
The Epstein law is valid for centimeter to meter-size dust particles at $\sim 5$AU 
if gas column density is less than twice as much as that of 
in the minimum solar nebula model \citep[MMSN;][]{hayashi81}.
Even if we consider higher column density, the deviation in the drag force strength 
from Epstein law would not be significant.
The last term in Equation~(\ref{eq:particle}) is the self-gravitational force of the dust particles. 
We calculate the gravitational potential $\Phi$ from the interpolated dust spatial density $\rho_d$,
 solving the Poisson equation, 
\begin{eqnarray}
{\nabla}^2 \Phi=4\pi G{\rho}_d. 
\label{eq:poisson}
\end{eqnarray}
We calculate the self-gravity of the dust particles only in the situations where the GI is expected. 
We neglect the vertical gravity of the host star that causes settling of dust particles 
onto a thin layer \citep[$\sim 0.01H$ where $H$ is the disk scale height;][]{schrapler04} near the midplane,
because we are interested in radial concentration of dust but not vertical settling.
While this would not affect the growth of MRI around the disk mid-plane that we simulate ($|z|<0.25H$), 
we do not have to resolve the thin dust layer, avoiding expensive computational cost. 

For the disk gas, we use the isothermal resistive MHD equations, 
\begin{eqnarray}
\frac{\partial \mathbf{u}}{\partial t}
 +\left(\mathbf{u}\cdot\nabla\right)\mathbf{u}&=&
 -\frac{1}{\rho_g}\nabla\left(P+\frac{\mathbf{B}^2}{8\pi}\right)
 +\frac{1}{4\pi\rho_g}\left(\mathbf{B}\cdot\nabla\right)\mathbf{B}
 -2\mathbf{\Omega}\times\mathbf{u}
 +3\Omega^2x\hat{\mathbf{x}} \nonumber\\
&& -\beta c_{s}\Omega\hat{\mathbf{x}}-\frac{\epsilon}{\tau_f}
\left(\mathbf{u}-\langle \mathbf{v} \rangle\right),
 \label{eq:motion}\\
\frac{\partial \rho_g}{\partial t}
 +\nabla\cdot\left(\rho_g \mathbf{u}\right)&=&0,\\
\frac{\partial \mathbf{B}}{\partial t} &=&
 \nabla\times\left[\left(\mathbf{u}\times\mathbf{B}\right)-
 \eta\left(\nabla\times\mathbf{B}\right) \right], \label{eq:induction}\\
P&=&c^2_{s}\rho_g,
\end{eqnarray}
where we assume constant $c_s$ (an isothermal disk), 
$\langle \mathbf{v} \rangle$ is the mean velocity 
field of the dust particles in the grid cell, $\epsilon=\rho_d/\rho_g$ is the dust to gas ratio,
and the term, $-\beta c_{s}\Omega\hat{\mathbf{x}}$, 
expresses the global pressure gradient, which is separated from the local one.
Let $P_0$ and $\delta P$ be the global pressure and the local pressure 
variation due to the effect of MRI ($P = P_0 + \delta P$).
Assuming that $P_0 \propto r^q$,  the global pressure gradient is given by 
\begin{eqnarray}
-\frac{1}{\rho_g}\frac{\partial P_0}{\partial r} = -\frac{1}{\rho_g} \frac{P_0}{r} q 
= -\frac{H}{r} q c_{s}\Omega 
= -\beta c_{s} \Omega,
\end{eqnarray}
where $H$ is the disk scale height defined by $H = c_s/\Omega$. 
In our local model, $\beta = qH/r$ is approximated to be constant.
Note that both $q$ and $\beta$ are negative.
Due to the radial pressure gradient, the gas rotation angular 
velocity is deviated from Keplerian one, as 
\begin{equation}
\Omega_g \simeq \Omega \left( 1 + \frac{1}{2}\left[\frac{H}{r}\right]^2 \frac{d \ln  P}{d \ln r}\right)=
\Omega \left( 1 + \frac{1}{2}\frac{H}{r} \beta + 
\frac{1}{2}\left[\frac{H}{r}\right]^2 \frac{d \ln \delta P}{d \ln r}\right).
\label{eq:p}
\end{equation}
The last term in the r.~h.~s. of Eq.~(\ref{eq:motion}) is 
the dust drag force (back-reaction of gas drag force on the dust particles).
Except for this term, the equations of motions for disk gas and dust particles
are the same as those in Paper II.
The purpose of this paper is to study the effect of this term on dust concentration. 
When dust particles are accumulated and $\epsilon$ takes a large value
of $\ga O(1)$, this term would influence the gas flow.
The induction equation (Eq.~[\ref{eq:induction}]) has 
the diffusion term by ohmic dissipation. 
Since we are interested in relatively inner regions  ($\lesssim$ a few dozens $\rm{AU}$),
we neglect the ambipolar diffusion that could influence MRI growth in the outer regions
of $\gtrsim 100{\rm AU}$ \citep{chi07}.
We treat the magnetic resistivity as a constant parameter for simplicity,
though in reality it depends on the density of tiny grains ($\lesssim \mu \rm m$). 

We scale length, time, and velocity by $H$, $1/\Omega$, and $c_s$ in simulations.
We solve the MHD equations by combining CIP scheme \citep{yabe91} and MOC-CT method \citep{stone92}.
The dust density is allocated to the closest eight grids in the three-dimensional space using cloud-in-cell (CIC) model. 
This algorithm strictly conserves angular momentum transfer from dust to gas by
using a similar method as \citet{johan07}.
The boundary conditions in all directions are periodic. 
For the radial boundary, however, we take into account
Keplerian differential rotation with the shearing box model \citep{wisdom88, haw95}. 
While we are not interested in vertical sedimentation, we want to keep total dust 
mass in a whole simulation area, so that we adopt the periodic boundary condition
also for vertical direction.
The Poisson equation~(\ref{eq:poisson}) is calculated by Fast Fourier Transform (FFT). 
This method requires the periodic boundary.
For radial direction, according to the sheared boundary,
we shift the phase azimuthally in Fourier space after the Fourier transform
in the periodic azimuthal direction, following \citet{johan07}. 
Our simulations are performed by a vector computer, NEC SX-9 at ISAS/JAXA. 

\subsection{Initial conditions}

The setup for the simulation in this paper is the same as CASE2 in Paper II. 
We assume non-uniform $B_z$ 
to set marginal MRI state where localized dead (stable) and active (unstable) regions co-exist 
in the initial conditions. 
The initial magnetic field is $\mathbf{B}_0=(0, B_{0}\sin\theta, B_{0}\cos\theta)$, where $\theta=\theta(x)$ is the angle between the magnetic field and the vertical axis. 
We assume a constant value of $B_0$ that is determined by 
plasma beta $=400$, to establish the initial equilibration. 
With the constant magnetic resistivity 
of $\eta=0.002H^2\Omega$ we adopt, a threshold vertical magnetic field for MRI is 
$B_{z,\rm crit}\sim 0.2B_0$ \citep{jin96}. 
We set radially varying $\theta$ 
in which $\theta=0^{\circ}$ ($B_z \gg B_{z,\rm crit}$) in the central 
zone and $\theta=85^{\circ}$ ($B_z<B_{z,\rm crit}$) in the side regions
(see Figure 2b in Paper II), such that MRI grows only in the central zone.  
We set the radial width of the initially active region as $L_{\rm u}=1.4H$ in all cases, 
and that of the dead regions as 
$L_{\rm s}=4.0H$ in model-s40-* and $L_{\rm s}=0.5H$ in model-s05-*,
where the asterisk * represents other simulation parameters (see below).

We assume equal-size dust particles
with $\tau_f=1.0/\Omega$ in model-s*-t10-* 
except for $\tau_f=0.1/\Omega$
in model-s40-t01-e010, neglecting their coalescence and 
fragmentation. 
At 5AU in MMSN, $\tau_{f}\Omega=0.1$ and $1.0$ correspond to the dust sizes of approximately 3 and 30 centimeters, respectively. 
The super-particles are distributed uniformly such that 
the initial dust-to-gas ratio $\epsilon_0 = 0.1$ for model-s*-t*-e010
and  $\epsilon_0=0.01$ for model-s40-t10-e001. 
Note that the dust-to-gas ratio in our simulation box 
corresponds to that near the midplane layer.
If vertically global sedimentation of dust particles is taken into account,
the dust-to-gas ratio in our simulation box should be
larger than that averaged over the whole disk.
For comparison, we also present the results of Paper II without
the dust drag as model-*-*-test.
The self-gravity of dust particles in Equation~(\ref{eq:particle}) are switched on 
only in the saturated state in which the GI is expected. 

All of our simulations start with uniform gas density and pressure. 
The global pressure gradient is set to be $\beta=-0.04$. 
This assumed value of $| \beta |$ is a few times smaller than that expected at $\sim 5$AU
in MMSN.
In order to compare the results with \citet{johan07} and Paper II,
we adopt the small value.
As was argued in Paper II, the results would not be affected significantly by
the value of $| \beta |$.
The initial angular velocities of gas and particles are 
$u_y/c_s = -\left(3/2\right)(x/H)+\beta/2$ and 
$v_y=-\left(3/2\right)(x/H)$, respectively. 
Because gas rotates slower, the particles migrate inward (negative direction of $x$) 
in the initial state. 
Initial disturbances are given randomly to the gas radial velocity 
with the amplitude of $0.001c_s$. 
The size of our simulation box is $(L_x,L_y,L_z)=((2.5$-$9.5)H, 1.0H, 0.5H)$ and the resolution is $dx=dy=dz=0.01H$. 
Eight particles are distributed in each grid initially and the total number is $\sim O(10^7)$. 
We have tested different number of distributed particles with different mass such that
total mass is conserved
and found that the results are converged if the distributed number of particles
for each grid is $\ga 8$. 
The initial setup is summarized in Table~\ref{tab:initial}.

\section{Effect of the dust drag on dust concentration}\label{sec:drag}

In the simulations with dust drag, quasi-steady state is formed 
and the dust particles are concentrated around the outer-edge of the super-Keplerian region 
by inhomogeneous MRI. 
Here, we study the effect of the dust drag on the dust concentration 
by comparing the results with those without the dust drag. 

\subsection{$\tau_f\Omega=1.0$ in a weak remnant turbulence}

In this subsection, we discuss the results with $\tau_f\Omega=1.0$,
model-s40-t10-e010 with $\epsilon_0 = 0.10$ and model-s40-t10-e001 with $\epsilon_0 = 0.01$.
Strong dust concentration was found in corresponding models
without the dust drag in Paper II¡¡ (model-s40-t10-test), which is summarized in Figure~\ref{fig:paper2}. 
MRI is excited only in the initially active central region ($-0.71 < x/H < 0.71$).
The MRI turbulence transfers mass and angular momentum of disk gas to establish
a local rigid rotation in the initially active zone through the turbulent viscosity.
Then, in the outer half region of the active zone, gas flow is
accelerated and migrates outward because of angular momentum gain.
On the other hand, in the inner half region, it is decelerated and migrates inward.
As a result, gas mass is moved from the central zone to the side zones
and gas pressure is lowered in the central zone.
The effect of the pressure modulation extends by radial scale of $\sim H$.
Equation~(\ref{eq:p}) shows that
\begin{equation}
\delta \tilde{u}_y = \frac{u_y - v_{\rm kep}}{c_s} \simeq 
\frac{\beta}{2}+ \frac{1}{2}\frac{H}{r} \frac{d \ln  \delta P}{d \ln r} =
- 0.02 +  \frac{1}{2}\frac{d \ln \delta P}{d \ln (x/H)}.
\end{equation}
This equation shows that 
super-Keplerian regions are associated with locally positive pressure gradient
with some off-set due to global pressure gradient.
We find that super-Keplerian regions are created in $0\lesssim x/H \lesssim 2.0$ and 
$-2.8 \lesssim x/H \lesssim -2.5$ (panel a and b).
Although MRI turbulence has decayed in the result at $t\Omega=70$,
the magnetic field has not been dissipated in the central zone (panel c).
The MRI is suppressed by the disappearance of shear motion.
If the rigid rotation is perturbed toward the original Keplerian motion,
the retrieved shear motion causes MRI again to recover the rigid rotation.
Thus, this profile is stable and strong.

\begin{figure}
\figurenum{2}
\epsscale{0.9}
\plotone{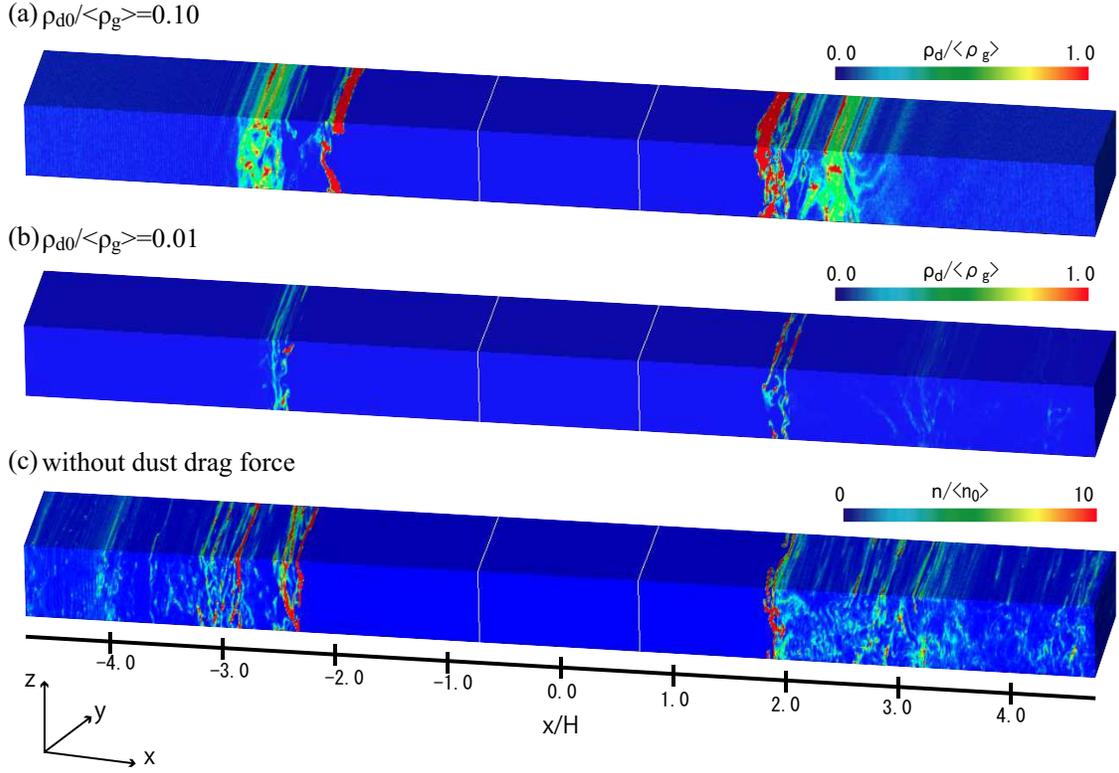}
\caption{
The dust density at the saturated state in  
(a) model-Ls40-t10-e010 ($\tau_f\Omega=1.0$ and $\epsilon_0=0.10$),  
(b) model-Ls40-t10-e001 ($\tau_f\Omega=1.0$ and $\epsilon_0=0.01$) and 
(c) model-Ls40-t10-test ($\tau_f\Omega=1.0$ without dust drag force). 
The sampling time is $t\Omega=104$. 
The initially active region is located between the two white lines. 
}
\label{fig:s40_contour}
\end{figure}

Figure~\ref{fig:s40_contour} presents the dust density in the saturated state 
in model-s40-t10-e010 with $\epsilon_0 = 0.10$ (panel a) 
and model-s40-t10-e001 with $\epsilon_0 = 0.01$ (panel b),
in comparison with model-s40-t10-test without the dust drag (panel c).
Because the dust drag depends on spatial density of the dust particles, 
the results depend on $\epsilon_0$.
In all cases, after the particles are swept out of the active region by the MRI turbulence, 
they accumulate at the outer-edge of the super-Keplerian zone at $x/H\simeq$ 2.0 and -2.5.
Particles leaving the simulation box from the small $x$ (left hand) boundary reenter 
the simulation region from the large $x$ (right hand) boundary after the shearing box correction is taken into account. 
The dust that reentered from the right hand boundary is halted at $x/H\simeq 2.0$, 
resulting in further increasing of the dust density. 
The number of locations of dust concentration is fewer in the case with the dust drag,
because the drag smoothes out small amplitude fluctuations of gas pressure.
In the case with the drag, the individual dust concentrated areas are 
broader in model-s40-t10-e010 than in model-s40-t10-e001,
which is discussed below.
We also found that velocity dispersion is lower for a denser clump, 
which was not observed in the case without the dust drag.
The effect of the reduced velocity dispersion will be discussed in section 4.

\begin{figure}
\figurenum{3}
\epsscale{0.6}
\plotone{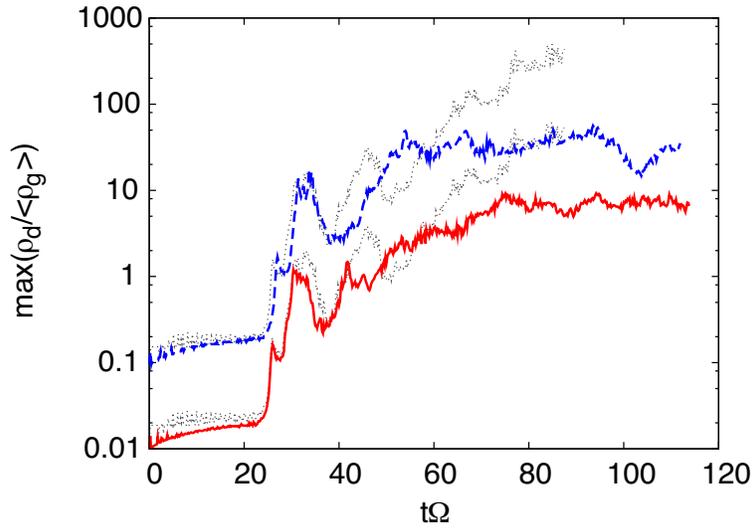}
\caption{
Time evolution of the maximum dust concentration 
in model-Ls40-t10-e010, -e001 and -test. 
The solid and dashed lines represent the results of $\epsilon_0=0.10$ 
and $\epsilon_0=0.01$, respectively. 
All lines represent the dust density in the cell having the highest density in the whole simulation region, 
which is normalized by the gas density averaged over the whole region ($\langle \rho_g \rangle$). 
The thin dotted lines show the result without the dust drag (model-Ls40-t10-test).
In this result, only concentration relative to the initial state is measured, so
these lines are drawn by assuming $\epsilon_0=0.10$ or $\epsilon_0=0.01$. 
}
\label{fig:s40_maxd}
\end{figure}

Figure~\ref{fig:s40_maxd} shows the time evolution of the maximum 
density of dust particles ($\rho_d$) in the simulation cells.
The dust density is scaled by the gas density averaged in the whole 
simulation region ($\langle \rho_g \rangle$). 
The results are compared with those without the dust drag 
(dashed lines; model-Ls40-t10-test) for
$\epsilon_0=0.10$ and $\epsilon_0=0.01$.
In the case without the dust drag, only concentration relative to the initial state is measured, so
these lines are drawn by the evolution of concentration 
assuming $\epsilon_0=0.10$ or $\epsilon_0=0.01$. 
The maximum density continues to increase monotonically in this case.
However, the growth of the maximum value is saturated in both cases with the dust drag.  
The saturation is faster and the increase rate is slightly smaller in model-s40-t10-e010
than in model-s40-t10-e001. 

\begin{figure}
\figurenum{4}
\epsscale{1.0}
\plotone{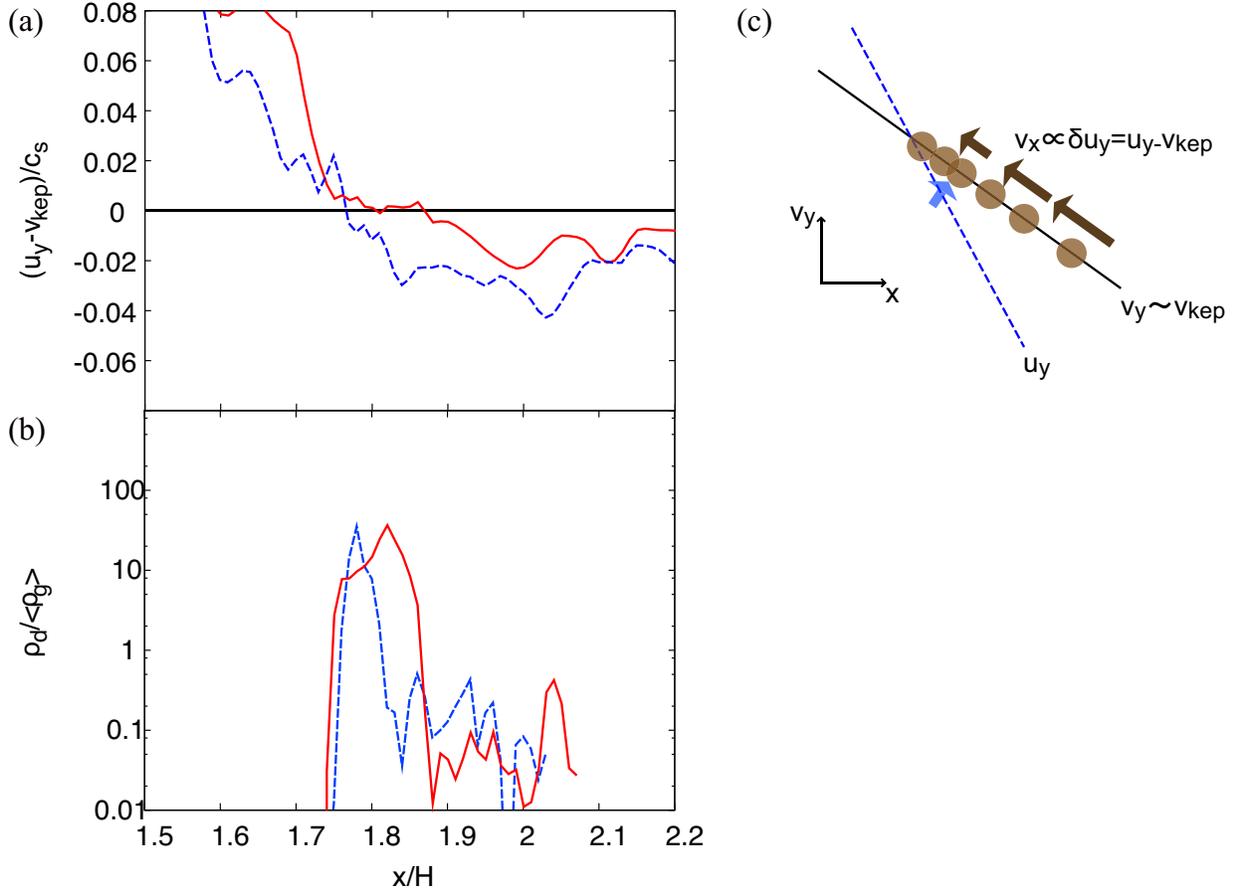}
\caption{
Broadening of the dust concentrated region by the dust drag.
(a) The difference between the gas angular velocity and Kepler angular velocity and 
(b) the vertically averaged dust density  
at $t\Omega=55.4$ (dashed lines) and $t\Omega=70.0$ (solid lines)
in model-Ls40-t10-e010 ($\tau_f\Omega=1.0$ 
and $\epsilon_0=0.10$). 
These figures are magnification of the area around the concentrated region where $u_y = v_{\rm kep}$. 
(c) Schematic illustration of "traffic jam" of the dust particles.
}
\label{fig:s40_expand}
\end{figure}

In order to explain the broadening of the dust accumulated region,
in addition to the three-dimensional simulations, we performed 
the two-dimensional ($x$-$z$) version of the model-s40-t10-e010,
in which the effect of the dust drag on the radial migration of the dust particles 
is more clearly shown.
In Figure~\ref{fig:s40_expand}a, we plot the difference of gas angular velocity from Kepler angular velocity, $\delta \tilde{u}_y=\left(u_y-v_{\rm kep}\right)/c_s$, near the sub- and super-Keplerian boundary
in the two-dimensional simulation.  
The velocities are averaged over the vertical direction. 
At $t\Omega=55.0$ (dashed line), the dust particles are expected to assemble 
at $x/H\simeq 1.75$, where $\delta \tilde{u}_y=0$. 
At $t\Omega=87.4$ (solid line), however,
the region with $\delta \tilde{u}_y\sim 0$ becomes broader ($1.7\lesssim x/H\lesssim 1.9$),
because more dust particles have migrated to this region and their drag
makes the gas flow close to Keplerian. 
Consequently, the dust particles are more broadly distributed there
(Figure~\ref{fig:s40_expand}b).
Figure~\ref{fig:s40_expand}c schematically illustrates the effect of
the dust drag. 
The radial velocity ($\propto \delta \tilde{u}_y$) of a migrating dust particle 
becomes slower as the dust particle approaches the dust-concentrated 
region of $\delta \tilde{u}_y=0$, like "traffic jam."
Due to the modulation by the dust drag,
the radial width of the Keplerian region is expanded, 
and the dust particles stop their inward migration
before they reach the location at which $\delta \tilde{u}_y=0$ originally.
Thus, the maximum dust density is self-regulated as shown in Figure~\ref{fig:s40_maxd}. 

In the three-dimensional simulation, similar results are obtained,
although small amplitude oscillations remain.
The similarity implies that geometry  is not the main cause for 
the broadening of the dust concentration region.

\subsection{$\tau_f\Omega=0.1$ in a weak remnant turbulence}

In Paper II, for smaller particles with $\tau_f \Omega=0.1$ (model-s40-t01-test), 
we found that the dust concentration is not significant because the smaller dust particles 
are affected more by the turbulent diffusion.
In the results in section 3.1, we found that the dust drag suppresses  
the local turbulence and velocity dispersion of the dust particles.
It could enhance the dust accumulation by decaying the turbulence around 
the dust accumulated area, as found in \citet{johan07}.

\begin{figure}
\figurenum{5}
\epsscale{0.6}
\plotone{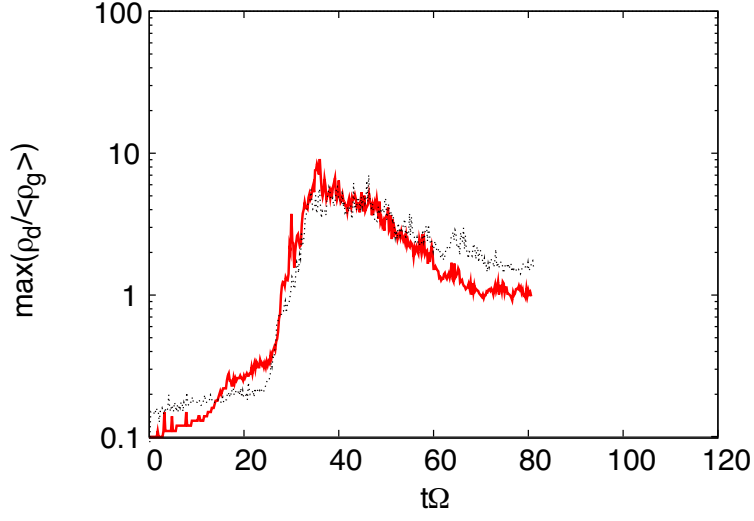}
\caption{
Same as Figure~\ref{fig:s40_maxd} but for model-s40-t01-e010 (bold solid line; $\tau_f\Omega=0.1$ and $\epsilon_0=0.10$) 
and -test (thin dotted line; $\tau_f\Omega=0.1$ without the dust drag force). 
}
\label{fig:s40t01_maxd}
\end{figure}

However, the time evolution of the maximum scaled density of dust particles 
($\rho_d/\langle \rho_g \rangle$) 
in model-s40-t01-e010 is not significantly different from that in model-s40-t01-test
(Figure~\ref{fig:s40t01_maxd}). 
In both cases, after $\rho_d$ rapidly increases by the diffusing-out 
from the initially active region by the MRI turbulence $t \Omega \sim 25$, 
it gradually increases and is saturated at $t \Omega \ga 50-70$.
The velocity dispersion of the dust particles is actually reduced from
that in the case without the dust drag, 
but the effect is not strong enough to enhance the dust concentration.

\subsection{$\tau_f\Omega=1.0$ in a strong remnant turbulence}

In Paper II, we found that 
with the smaller initial dead region $L_{\rm s}=0.55H$, the viscosity 
in the saturated state is $\alpha\sim 10^{-2}$, which is much larger than 
$\alpha \sim 10^{-4}$ for the runs with
$L_{\rm s}=4.0H$ in model-Ls40-*. 
We found in Paper I that MRI turbulence does not
decay if the magnetic
Elssaser number radially averaged over the simulation region
is smaller than unity in the initial state.
Elsasser number is defined by
\begin{eqnarray}
\Lambda_{\rm m,ave}=v^2_{{\rm A}z}/\eta\Omega, \label{Rmave}
\label{eq:lambda}
\end{eqnarray}
where $v_{{\rm A}z} = B_z /\sqrt{4 \pi \rho_g}$ is 
$z$ component of Alfven velocity and $\rho_g$ is spatial density of the disk gas. 
The run with $L_{\rm s}=0.55H$ corresponds to $\Lambda_{\rm m,ave} \sim 0.5$.

\begin{figure}
\figurenum{6}
\epsscale{0.6}
\plotone{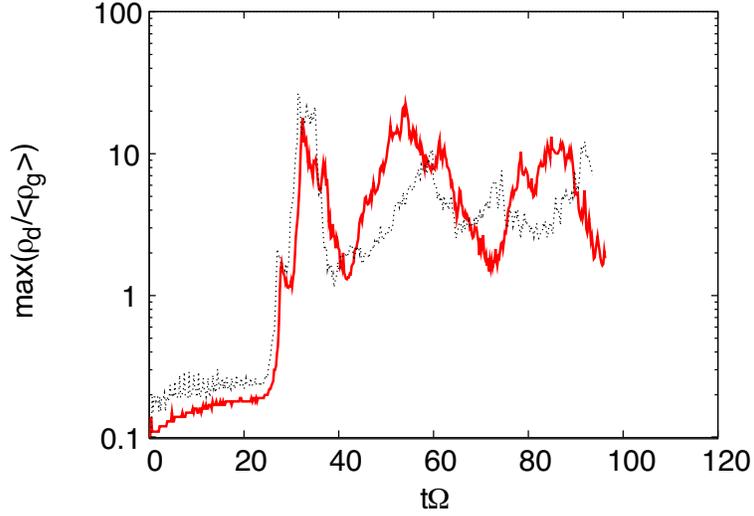}
\caption{Same as Figure~\ref{fig:s40_maxd} but for model-s05-t10-e010
(bold solid line; $L_{\rm s}=0.55H$) 
and -test (thin dotted line; without the dust drag force). 
}
\label{fig:s05_maxd}
\end{figure}

The stronger remnant turbulence limited $\rho_d/\langle \rho_g \rangle$
to the values less than 100 even for $\tau_f \Omega=1.0$ in the case without the dust drag.
We performed the run with $L_{\rm s}=0.55H$,
adding the dust drag (model-s05-t10-e010). 
Although the drag force reduces the turbulent diffusion 
in the local concentrated region, the maximum dust density
is not enhanced from that in the case without the dust drag
(Figure~\ref{fig:s05_maxd}). 

\section{Planetesimal formation}\label{sec:planetesimal}

As shown in the previous section, the dust accumulation is
self-regulated by the effect of the dust drag.
The suppressed maximum density of dust particles 
is unfavorable for the gravitational instability (GI), while
their reduced velocity dispersion is favorable.
In this section, we analyze the results of previous runs 
without the self-gravity of dust particles to examine
the possibility of subsequent GI.
In the results of some favorable runs, 
we re-perform the simulations, including self-gravity among 
the dust particles, to demonstrate that the GI actually occurs.

\subsection{Analysis of gravitational instability}

The GI is expected to arise when the self-gravity of dust particles is dominant over their thermal fluctuation (velocity dispersion), in other words, when the radius (size) $R$ of 
a dust clump with mass $M$ is smaller than the Jeans (Bondi) radius,
\begin{eqnarray}
R_{\rm J}=\frac{2GM}{\sigma^2}
=2\left(\frac{\sigma}{c_s}\right)^{-2}
\frac{M}{M_{\ast}} \left(\frac{H}{R}\right)^{-3} H,
\label{eq:RJ}
\end{eqnarray}
where $M = M(R)$ and $\sigma = \sigma(R)$ are the total mass 
and mean velocity dispersion of particles in the region with distance $R$ from the
center of the clump,
$M_{\ast}$ is the host star's mass, 
$c_s = \Omega H$, and $\Omega = \sqrt{GM_{\ast}/r^3}$.
The background shear is included in $\sigma$.
In order to numerically resolve GI, we set grid size in the $x$-direction 
such that $dx < 0.5 R_{\rm J}$. 

The condition of GI is often described by linear analysis of a uniform axisymmetric disk
\citep[e.g.,][]{safronov69, gold73,sekiya83}, which is essentially
equivalent to Toomre's condition \citep{toomre64},
\begin{equation}
1 < Q = \frac{\Omega \sigma}{\pi G \Sigma_d},
\end{equation}
where $\Sigma_d$ is the unperturbed solid column density.
If we use $M \sim \pi \Sigma_d R^2$ and $\sigma \sim R\Omega$,
the Toomre's condition is identical to $R < R_{\rm J}$ except a numerical factor of 2. 
Because significant radial inhomogeneity develops before GI occurs in our
case, we use the condition $R < R_{\rm J}$ that can be locally applied, rather than $Q < 1$.

The mass of each super-particle $m$ is given by $\rho_{d0}/n_0$,
where $\rho_{d0}$ and $n_0$ are the spatial density and 
number density of particles in the initial conditions.
Then, the dust clump mass is given by
\begin{eqnarray}
M = m N_R =\frac{N_R}{n_0}\rho_{d0}
     = \sqrt{\frac{8\pi^3}{9}} \frac{N_R}{N_{R0}} 
         \left(\frac{H}{r}\right)^{-2}
         \epsilon_0 \frac{\Sigma_{g0} r^2}{M_{\ast}}
         \left(\frac{R}{H}\right)^{3}
         M_{\ast}.
\label{eq:m_super_particle}
\end{eqnarray}
where $N_R$ is the total particle number in $R$,
$N_{R0}=(4\pi/3)n_0 R^3$ is its initial value,
$\rho_{d0} = \epsilon_0 \rho_{g0}$ and
$\rho_{g0} = \Sigma_{g0}/\sqrt{2\pi} H$ by the assumption of 
a vertically isothermal disk.
From Equtaions~(\ref{eq:RJ}) and (\ref{eq:m_super_particle}),
the scaled Jeans radius 
is given by
\begin{eqnarray}
\frac{R_{\rm J}}{H} &=& \sqrt{\frac{32\pi^3}{9}} 
\left(\frac{\sigma}{c_s}\right)^{-2} 
\frac{N_R}{N_{R0}} \left(\frac{H}{r}\right)^{-5} 
\epsilon_0 \frac{\Sigma_{g0} r^2}{M_\ast}
\left( \frac{R}{H} \right)^{3}, 
\label{eq:RJ2} 
\end{eqnarray}
The initial dust-to-gas density ratio $\epsilon_0$ is given.
The simulation results give dust enhancement in a clump $N_R/N_{R0}$, 
the clump size $R/H$, and velocity dispersion in the clump $\sigma/c_s$
(the equations of motions we use are normalized by $H$ and $c_s$).
From the results and simulation parameters,  
we can evaluate $R_{\rm J}$ 
for any given values of $\Sigma_{g0}r^2/M_\ast$ and $H/r$ that are
specified by disk model through Eq.~(\ref{eq:RJ2}).

\subsection{Effect of dust drag force on $R_{\rm J}$}

The onset of GI depends on $N_{R}$ and $\sigma$ as shown by Equation~(\ref{eq:RJ2}). 
The dust drag suppresses $N_{R}$ and reduces $\sigma$ (section~\ref{sec:drag}). 
Here, using the arguments in section 4.1., 
we examine the possibility of the GI in model-s40-t10-e010, which is
one of the most promising runs for the GI.

\begin{figure}
\figurenum{7}
\epsscale{0.6}
\plotone{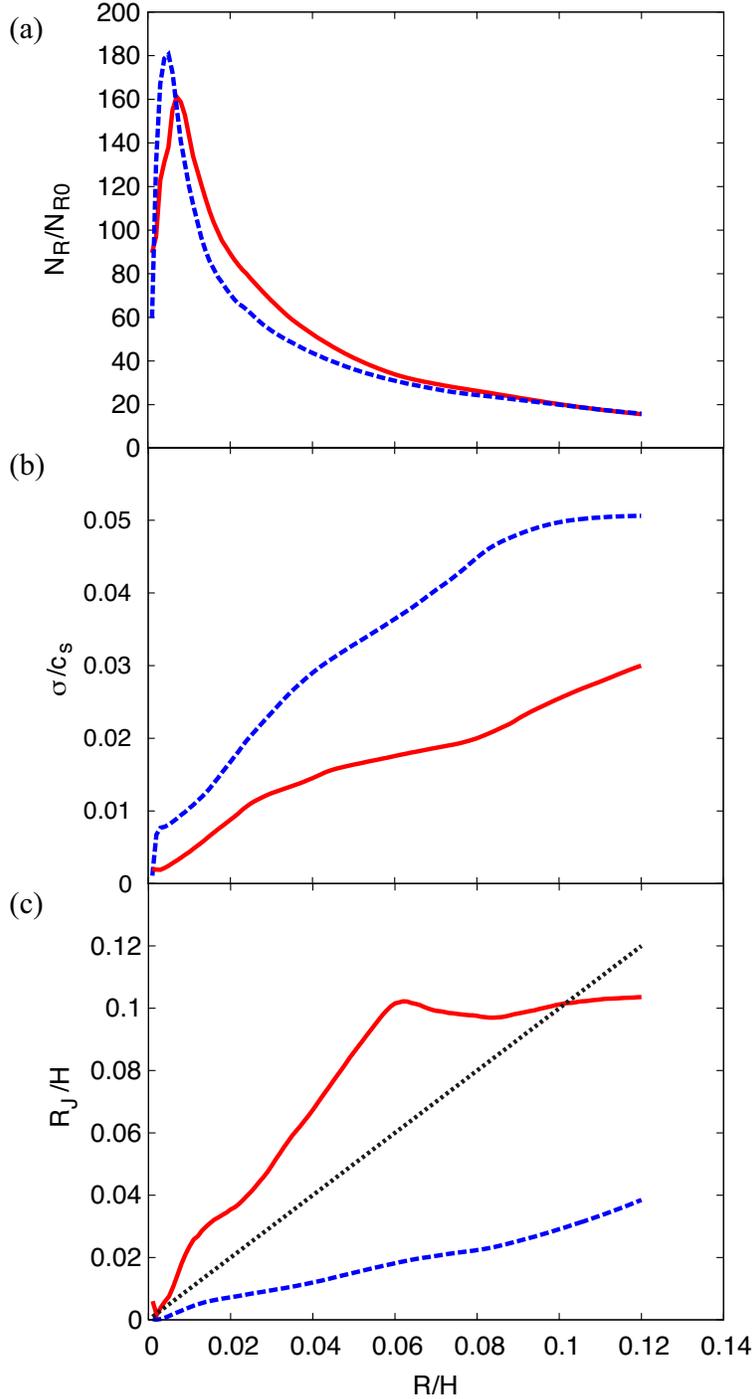}
\caption{Estimation of possibility of the GI by Eq.~(\ref{eq:RJ2}).  
In all panels, the solid and dashed lines represent model-s40-t10-e010 and model-s40-t10-test
at $t\Omega=58.0$, respectively. 
(a) The number $N_R$ of particles within distance $R$ from a densest grid
normalized by the initial value $N_{R0}$.
(b) The velocity dispersion of the particles in $R$. 
(c) The radius $R_{\rm J}$ calculated for each $R$ for
$M_\ast=M_{\odot}$, $r=5{\rm AU}$ and $\Sigma_{g0}=150 {\rm g cm^{-3}} \sim \Sigma_{\rm MMSN} (r=5{\rm AU})$.
In the region over the dotted line ($R_J > R$), the GI is expected. 
}
\label{fig:s40_Jeans}
\end{figure}

Figure~\ref{fig:s40_Jeans} shows  $N_{R}/N_{R0}$ 
from the densest grid point (panel a) and 
the corresponding $\sigma/c_s$ (panel b) in the results of
model-s40-t10-e010 and model-s40-t10-test at $t\Omega=58.0$.
Panel c shows Jeans radius $R_{\rm J}$ calculated 
for $M_\ast=M_{\odot}$, $r=5{\rm AU}$, $H/r=0.055$ and $\Sigma_{g0}=150 {\rm g cm^{-3}} (\sim \Sigma_{\rm MMSN} \; {\rm at} \; r=5{\rm AU}$, where $\Sigma_{\rm MMSN}$ is
gas column density in MMSN). 
While the particle concentration is lowered by the dust drag only slightly (panel a),
the velocity dispersion is significantly reduced (panel b).
Since the positive effect for the GI (reduction in the velocity dispersion)
is dominated over the negative one (suppression in particle concentration),
$R_{\rm J}$ calculated 
by Eq.~(\ref{eq:RJ2}) is higher and  
the condition for the GI ($R_{\rm J} > R$) 
is satisfied in the case with the dust drag (panel c).

\subsection{Simulation with dust self-gravity}

We carried out additional simulations including the self-gravity force of dust particles to demonstrate 
the formation of gravitationally bound clumps that may lead to planetesimals,
in model-s40-t10-e010. 
We set $M_\ast=M_{\odot}$, $r=5{\rm AU}$, $H/r=0.055$ and $\Sigma_{g0}=280 {\rm g cm^{-3}} \sim 2\Sigma_{\rm MMSN} (r=5{\rm AU})$. 
To reduce simulation cost, we introduced the self-gravity at $t\Omega=96$ 
when the dust concentration becomes saturated and $R_{\rm J}$ is 
much larger than the grid size $dx$.

\begin{figure}
\figurenum{8}
\epsscale{1.0}
\plotone{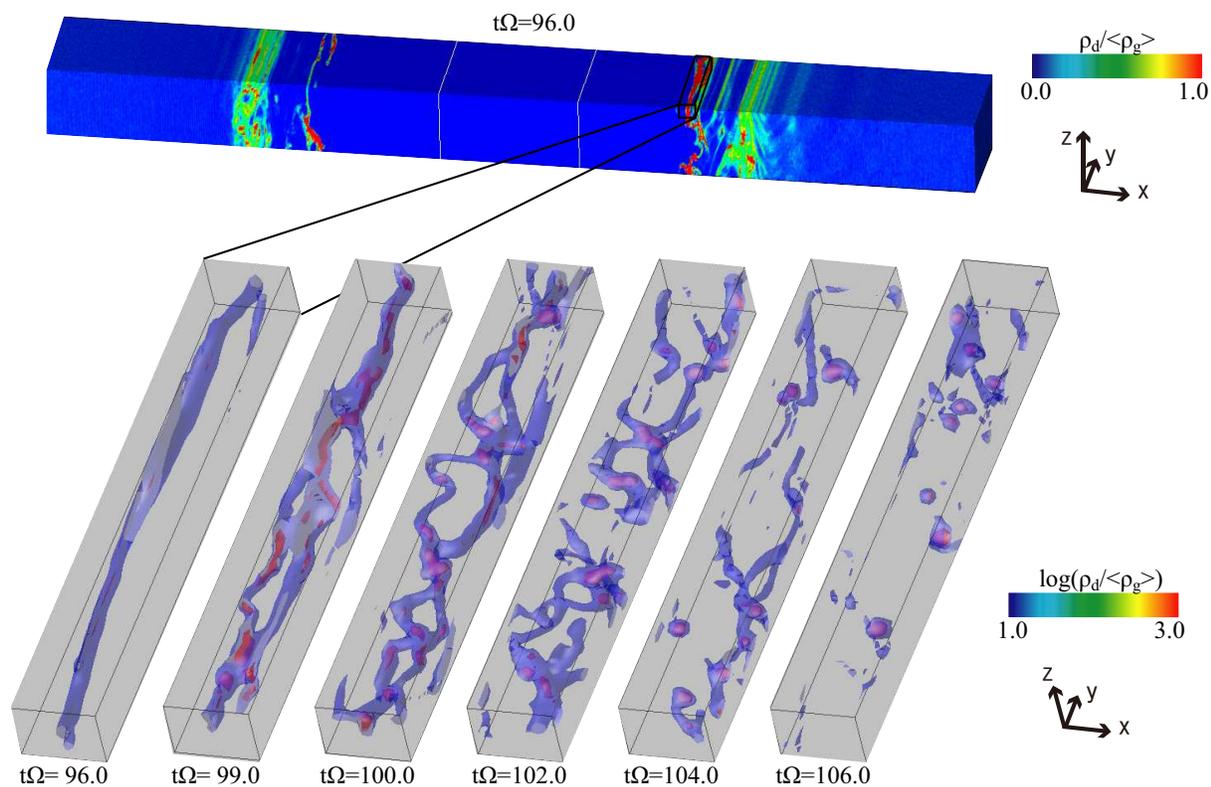}
\caption{Simulation of the GI with self-gravity of the dust particles
in model-Ls40-t10-e010. 
The top panel shows the dust density at $t\Omega=96.0$, at which 
the self-gravity of particles is added. 
The bottom panels show the time evolution of the GI. 
The different gray colors represent the isosurface of the dust-density 
$\log\left(\rho_d/\langle \rho_g \rangle \right)=1.0, 2.0$ and 3.0. 
The several clumps become bounded gravitationally. }
\label{fig:s40_self-iso}
\end{figure}

Figure~\ref{fig:s40_self-iso} shows the time evolution of the gravitational collapse. 
Shortly after introduction of the self-gravity, the elongated high density region 
is kinked ($t\Omega=99$) and it is separated into several clumps ($t\Omega=100$). 
The clumps grow by accreting surrounding dust particles and other clumps ($t\Omega=120$-140). 

\begin{figure}
\figurenum{9}
\epsscale{0.6}
\plotone{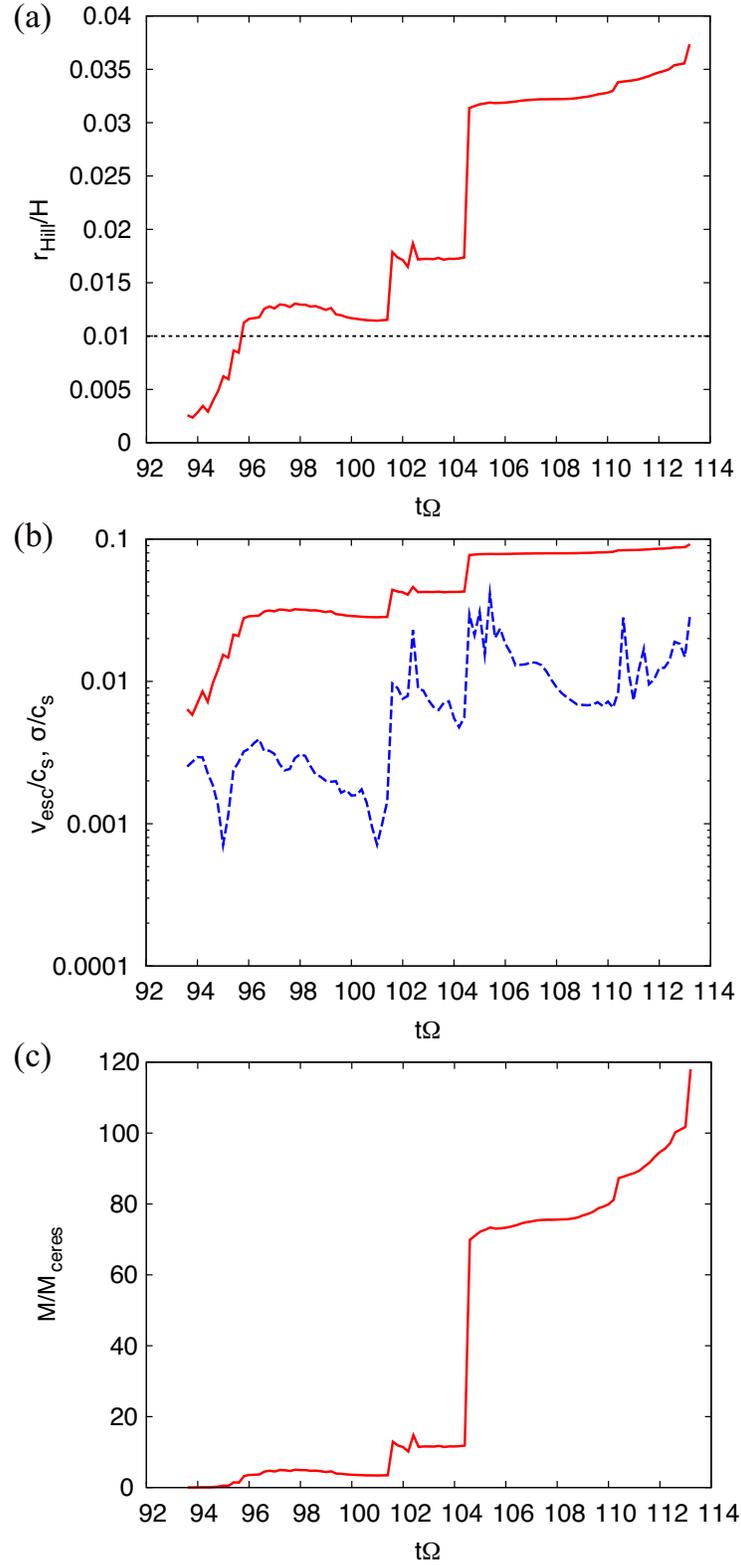}
\caption{Evolution of the gravitationally bounded clump in model-Ls40-t10-e010. (a) The Hill's radius,
(b) the surface escape velocity ($v_{\rm esc}$; the solid line) 
and the velocity dispersion of particles ($\sigma$; the dashed line) in the Hill's radius,
and (c) the total mass of the dust particles in the Hill's radius. }
\label{fig:s40_rH}
\end{figure}

To confirm that the clumps are gravitationally bound and 
estimate the mass of formed planetesimals, 
we define the range of a clump by its Hill's radius.
The time evolution of the clump is shown in Figure~\ref{fig:s40_rH}.
Figure~\ref{fig:s40_rH}a shows the Hill's radius, where the grid size is represented by a dashed line.
Immediately after the introduction of the self-gravity at $t\Omega=96$,
the Hill radius exceeds the grid size.
After that,  the clump is numerically resolved. 
Figure~\ref{fig:s40_rH}b shows that the velocity dispersion of the particles
in the clump is always
smaller than the surface escape velocity $v_{\rm esc}$ of the clump, which implies that
the clump is gravitationally bound.
If the gravitational collapse continues, it may form a planetesimal,
although this simulation does not have resolution to follow the subsequent collapse.

Figure~\ref{fig:s40_rH}c shows the temporal development of the clump mass
$M$, which may correspond to the planetesimal mass. 
The abrupt jumps at $t\Omega \sim 102$ and $\sim104$ are 
caused by collisional merging with other clumps. 
Since the destruction process is not properly included in our simulation, 
such rapid growth may be unrealistic. 
A conservative estimate for the planetesimal mass may be 
the mass before the abrupt jumps, that is, $\sim 4$ times Ceres mass. 
However, note that this mass is close to the resolution of our simulation
(Figure~\ref{fig:s40_rH}a) and the clump mass may be smaller 
in a higher-resolution simulation \citep{johan10}. 

We also performed the simulation with the self-gravity in model-s40-t10-e001, 
in which the Jeans radius is slightly larger than our grid size only for short interval. 
The GI is not found in this case, but it might be seen in a high-resolution simulation. 

\subsection{Critical gas column density for gravitational instability}

In the simulation with addition of the self-gravity in section 4.3, we assumed 
$\Sigma_{g0}$ that corresponds to $\sim 2\Sigma_{\rm MMSN}$ at $r=5{\rm AU}$. 
On the other hand, simulations before adding self-gravity 
are scaled by $\Sigma_{g0}r^2/M_*$ and $H/r$.

Here, fixing $r=5$AU and $H/r=0.055$, we apply the results of individual runs 
for various $\Sigma_{g0}$
to derive a sufficient condition for gas column density to cause the GI.
The conditions for the GI is $R<R_{\rm J}(R)$.
Since $R_{\rm J} \propto \Sigma_{g0}$  (Equation~[\ref{eq:RJ2}]),
the condition is more easily satisfied for larger $\Sigma_{g0}$. 
In the saturated state in model-s40-t10-e010, 
the condition is satisfied even at $\Sigma_{g0}/\Sigma_{\rm MMSN}\sim 1$,
while we showed the results with $\Sigma_{g0}=2\Sigma_{\rm MMSN}$ in section 4.3. 
For smaller $\epsilon_0$ (model-s40-t10-e001), 
the critical column density is $\Sigma_{g0}/\Sigma_{\rm MMSN} \sim 3$. 
The smaller dust particles with $\tau_f\Omega=0.1$ have no chance to 
excite the GI even in the weak residual turbulence (model-s40-t01-e010)
as long as $\Sigma_{g0}/\Sigma_{\rm MMSN}< 20$. 
In the stronger remnant turbulence ($L_{\rm s}=0.5H$), 
the GI is not expected unless $\Sigma_{g0}/\Sigma_{\rm MMSN}>10$ (model-s05-t10-e010). 

Note that $R_{\rm J}$ is a function of $\epsilon_0 \Sigma_{g0}$
(Eq.~[\ref{eq:RJ2}]).
That is, the possibility of the GI depends on the total column density of dust particles, 
but not on $\epsilon_0$. 
Thus, for example, $R_{\rm J}$ should be similar between the result
with $\Sigma_{g0}/\Sigma_{\rm MMSN}=1$ in model-Ls40-t10-e010 ($\epsilon_0=0.10$) 
and $\Sigma_{g0}/\Sigma_{\rm MMSN}=10$ in model-Ls40-t10-e001 ($\epsilon_0=0.01$)
at the same $r$.
From the results of simulations in this paper, 
it is inferred that the GI may occur when $\rho_{d0,\rm{crit}}\gtrsim 0.03 \rho_{g, \rm MMSN}$.

\section{Conclusion and Discussion}\label{sec:discussion}

We have studied the dust concentration including the "dust drag force" onto gas 
 (back-reaction of the gas drag exerted onto the dust particles) in a quasi-steady state created by inhomogeneous MRI found by Paper I and II,  
by performing the three-dimensional resistive MHD simulation including dust particles as super-particles.
Since the inertia of the particles is taken into account,
the dust drag force modulates gas flow in the dust concentrated regions. 
We also examined the possibility of the planetesimal formation via gravitational instability (GI) 
with analysis using Jeans radius of dense dust regions
and performed simulations with adding self-gravity of the dust particles
to demonstrate that gravitationally bound clumps are actually formed by the GI.

If MRI active and dead zones initially coexist,
mass and angular momentum transfer associated by non-uniformly growing MRI turbulence 
changes the slightly sub-Keplerian gas flow in the
initial state to a quasi-steady MRI-stable state in which super- and sub-Keplerian regions are 
radially adjacent to each other (Paper I),
and the dust particles are concentrated at the outer edge of the super-Keplerian region (Paper II). 
In this paper, we found that the introduction of the dust drag 
broadens the dust accumulated regions while it reduces
velocity dispersion of the particles, depending on the turbulent level
and the friction time of the dust particles. 
We found that the positive effect (the reduction in velocity dispersion)
is generally dominated over the negative effect (the broadening of 
the dust accumulated region). 
Consequently, in the case with dust drag, the GI is expected
in the case of weak remnant turbulence (the turbulent viscosity $\alpha \sim 10^{-4}$)
and meter-size particles with $\tau_f\Omega=1.0$, if the initial dust spatial density is a few times larger
than that of MMSN.
The GI is regulated by the absolute value of the dust spatial density, but not
by the dust-to-gas ratio.

Smaller dust particles ($\tau_f\Omega=0.1$) are also less likely to cause the GI
even in the weak remnant turbulence case,
because they are more strongly coupled with gas turbulent motion. 
Since dust particles should have size distribution, 
the spatial density contributed from meter-size particles 
must be a few times larger than total dust density of MMSN for the GI.
However, dust settling increases dust density in the layer of midplane 
that corresponds to our simulation box and 
it could compensate for the effect of size distribution.
If the vertical dust distribution is Gaussian ($\propto e^{-z^2/2H^2}$) before the dust settling
and it is assumed that most of dust particles settle down to our simulation box with $L_Z = 0.5H$,
the averaged dust-to-gas ratio of our simulation box
is enhanced by a factor of five from the initial dust-to-gas ratio of the whole disk. 

In the models for dust growth in turbulent eddies  
proposed by other authors,  
high collision velocity between the dust particles excited by the turbulence
may result in fragmentation rather than coalescence, which is not favored 
for planetesimal formation. 
However, in our model, MRI turbulence is almost terminated after it transforms
the gas flow to the quasi-steady state with the pressure bump,
so that the collision velocities between dust particles
are as small as $\lesssim 0.6$--$0.7{\rm m/s}$ at $r=3$--$5\rm{AU}$,
which can avoid fragmentation at mutual collisions.

Radially non-uniform excitation of MRI is an essential point for the emergence of 
the pressure bump in our model.
The growth rate of MRI depends on the strength of the magnetic field and resistivity.
In Paper II, we found non-uniform resistivity produces 
the same quasi-steady state as non-uniform magnetic field that we 
assume in this paper.
In section 1, we raised a possibility of formation of the active zone
radially sandwiched by dead zones due to non-uniform resistivity near the snow line.
However, the location of the snow line and dead zones are
coupled with disk evolution due to viscous diffusion and photoevaporation
and also with growth, fragmentation and migration of dust particles.
Thus, to evaluate the possibility of radially "local" planetesimal formation proposed by
this paper, 
full-scale coupled evolution of the snow line, the dead zone, the disk gas ionization degree and dust growth needs to be studied theoretically and
by observation with ALMA. 

%

\acknowledgments
We thank for detailed comments by an anonymous referee.
This work was supported by Grant-in-Aid for JSPS Fellows (208778). The simulations presented in this paper were performed by NEC SX-6 at ISAS/JAXA.


\clearpage

\clearpage
\clearpage
\clearpage

\clearpage
%
%
\clearpage
\clearpage
\clearpage
\clearpage
\begin{table}
\begin{center}
\begin{tabular}{rcccc}
\hline \hline
Run & $L_{s}$ & $\tau_f\Omega$ & $\epsilon_0$ & self-gravity \\
\hline

model-s40-t10-e010 & 4.0$H$ & 1.0 & 0.10 & off and on \\
                 -t10-e001 & 4.0$H$ & 1.0 & 0.01 & off and on \\
                 -t10-test$\;$ & 4.0$H$ & 1.0 & 0.0 & off \\
                 -t01-e010 & 4.0$H$ & 0.1 & 0.10 & off \\
                 -t01-test$\;$ & 4.0$H$ & 0.1 & 0.0 & off \\
model-s05-t10-e010 & 0.55$H$ & 1.0 & 0.10 & off \\
                 -t10-test$\;$ & 0.55$H$ & 1.0 & 0.0 & off \\
\hline
\end{tabular}
\end{center}
\caption{Setup of individual runs.
$L_{\rm s}$ is the radial width of initially dead region;
$\tau_f$ is friction time; $\epsilon_0$ is initial dust-to-gas density ratio.
Model-s40-t10-e010 and e001 are also re-started with introduction of
self-gravity of particles. }
\label{tab:initial}
\end{table}


\begin{thebibliography}{}

\bibitem[Bai \& Stone(2010a)]{bai10a}
Bai, X.-N. \& Stone, J. M. 2010, \apj, 722, 143
\bibitem[Bai \& Stone(2010b)]{bai10b}
Bai, X.-N. \& Stone, J. M. 2010, \apj, 722, L220
\bibitem[Bai \& Stone(2010c)]{bai10c}
Bai, X.-N. \& Stone, J. M. 2010, \apjs, 190, 297
\bibitem[Balbus \& Hawley(1998)]{balbus98}
 Balbus, S. A., \& Hawley, J. F. 1998, Reviews of Modern Physics, 70, 1
\bibitem[Barge \& Sommeria(1995)]{barge95}
Barge, P., \& Sommeria, J. 1996, \aap, 295, 1
\bibitem[Barranco(2009)]{barranco09}
Barranco, J. 2009, \apj, 691, 907
\bibitem[Bulm and Wurm(2008)]{blum08}
 Bulm, J., \& Wurm, G. 2008, \araa, 46, 21
\bibitem[Brauer et al.(2008)]{brauer08}
 Brauer, F., Henning, Th., \& Dullemond, C. P. 2008, \aap, 487, L1
\bibitem[Chavanis (2000)]{chavanis00}
 Chavanis, P. H. 2000, \aap, 356, 1089
\bibitem[Chiang(2008)]{chiang08}
 Chiang, E. I. 2009, \apj, 675, 1549
\bibitem[Chiang \& Murray-Clay(2007)]{chi07}
 Chiang, E. I., \& Murray-Cley, R. A. 2007, Nature Phys., 3, 604
 \bibitem[Cuzzi et al.(1993)]{cuzzi93}
 Cuzzi, J. N., Dobrovolskis, A. R. \& Champney, J. M. 1993, Icarus, 106, 102
 \bibitem[Cuzzi \& Zahnle(2008)]{cuzzi08}
 Cuzzi, J. N., \& Zahnle, K. J. 2008, \apj, 614, 490
\bibitem[Dominik \& Tielens(1997)]{dominik97}
 Dominik, C., \& Tielens, A. G. G. M. 1997, \apj, 480, 647
\bibitem[Dzyurkevich et al.(2010)]{dzy10}
 Dzyurkevich, N., Frock, M., Turner, N. J., Klahr, H., \& Henning, Th. 2010, \aap, 515, A70
\bibitem[Fleming et al.(2000)]{fleming00}
 Fleming, T. P., Stone, J. M., \& Hawley, J. F. 2000, \apj, 530, 464
\bibitem[Fromang \& Papaloizou(2007)]{fromang07a}
 Fromang, S., \& Papaloizou, J. 2007, \aap, 476, 1113
\bibitem[Gammie(1996)]{gammie96}
 Gammie, C. F. 1996, \apj, 457, 355
\bibitem[Goldreich \& Ward(1973)]{gold73}
 Goldreich, P., \& Ward, W. R. 1973, \apj, 183, 1051
\bibitem[G\"{u}ttler et al.(2009)]{guttler09}
 G\"{u}ttler, C., Blum, J., Zsom, A., Ormel, C. W., \& Dullemond, C. P. 2009, \aap, 513, A56
\bibitem[Hayashi(1981)]{hayashi81}
 Hayashi, C., 1981, Prog. Theor. Phys. Suppl. 70, 35
\bibitem[Hawley et al.(1995)]{haw95}
 Hawley, J. F., Gammie, C. F., \& Balbus, S. A. 1995, \apj, 440, 742
\bibitem[Ida \& Lin(2004)]{ida04}
 Ida, S., \& Lin, D. N. C. 2004, \apj, 604, 388
\bibitem[Ida \& Lin(2008)]{ida08}
 Ida, S., \& Lin, D. N. C. 2008, \apj, 673, 487
\bibitem[Igea \& Glassgold(1999)]{igea99}
 Igea, J., \& Glassgold, A. E. 1999, \apj, 518, 848
\bibitem[Inaba \& Barge(2006)]{inaba06}
 Inaba, S., \& Barge, P. 2006, \apj, 649,415
 \bibitem[Ishitsu \& Sekiya(2003)]{ishitsu03}
 Sekiya, M. \& Ishitsu, N. 2003, Icarus, 165, 181
\bibitem[Jin(1996)]{jin96}
 Jin, L. 1996, \apj, 457, 798
\bibitem[Johansen et al.(2004)]{johan04}
 Johansen, A., Andersen, A. C., \& Brandenburg, A. 2004, \aap, 417, 361
\bibitem[Johansen et al.(2006a)]{johan06}
 Johansen, A., Klahr, H., \& Henning, Th. 2006, \apj, 636, 1121
\bibitem[Johansen et al.(2007)]{johan07}
 Johansen, A., Oishi, J. S., Mac Low, M,-M., Klahr, H., Henning, Th., \& Youdin, A. 2007, \nat, 448, 1022
\bibitem[Johansen \& Youdin(2007)]{johan07:si}
 Johansen, A., \& Youdin, A. N. 2007, \apj, 662, 627
\bibitem[Johansen et al. (2009)]{johan}
 Johansen, A., Youdin, A., \& Low, M. 2009, \apj, 704, L75
\bibitem[Johansen et al.(2010)]{johan10}
 Johansen, A., Klarhr, H., \& Henning, Th. 2010, arXiv1010.4757J
\bibitem[Kato et al.(2009)]{kato09}
 Kato, M. T., Nakamura, K., Tandokoro, R., Fujimoto, M., \& Ida, S.  2009, \apj, 691, 1697 (Paper~I)
\bibitem[Kato et al.(2010)]{kato10}
 Kato, M. T.,  Fujimoto, M., \& Ida, S. 2010, \apj, 714, 1155 (Paper~II)
\bibitem[Kokubo \& Ida(1998)]{kokubo98}
 Kokubo, E., \& Ida, S. 1998, Icarus, 77, 330
\bibitem[Kretke \& Lin(2007)]{kretke07}
 Kretke, K. A., \& Lin, D. N. C. 2007, \apj, 664, L55
\bibitem[Lee et al.(2010a)]{lee10a}
Lee, A. T., Chiang, E., Asay-Davis, X. \& Barranco, J. 2010, \apj 718, 1367L
\bibitem[Lee et al.(2010b)]{lee10b}
Lee, A. T., Chiang, E., Asay-Davis, X. \& Barranco, J. 2010, \apj 725, 1937L
\bibitem[Malhotra(1993)]{malhotra93}
 Malhotra, R. 1993, \nat, 365, 819
\bibitem[Morbidelli et al.(2009)]{morbidelli09}
 Morbidelli, A., Bottke, W. F., Nesvorny, D., \& Levison, F. 2009, Icarus, 204, 558
\bibitem[Nakagawa et al.(1981)]{nakagawa81}
Nakagawa, Y., Nakazawa, K. \& Hayashi, C. 1981, Icarus, 45, 517
\bibitem[Safronov(1969)]{safronov69}
 Safronov, V. S. 1969, Evolution of the Protoplanetary Cloud
 and the Planets, NASA Tech. Transl. F-677
\bibitem[Saito \& Sirono(2011)]{saito11}
 Saito, E., \& Sirono, S. 2011, \apj, 728, 20
\bibitem[Sano et al.(1998)]{sano98}
 Sano, T., Inutsuka, S., \& Miyama, S. M. 1998, \apj, 506, L57
\bibitem[Sano \& Miyama(1999)]{sano99}
 Sano, T., \& Miyama,S. M. 1999, \apj, 515, 776
\bibitem[Sano et al.(2000)]{sano00}
 Sano, T., Miyama, S. M., Umebayashi, T., \& Nakano, T. 2000, \apj, 543, 486
\bibitem[Schr\"{a}pler \& Henning(2004)]{schrapler04}
 Schr\"{a}pler, R., \& Henning, Th. 2004, \apj, 614, 960
\bibitem[Sekiya(1983)]{sekiya83}
 Sekiya, M. 1983, Prg. Theor. Phys., 69, 1116
 \bibitem[Sekiya(1998)]{sekiya98}
 Sekiya, M. 1998, Icarus, 133, 298
\bibitem[Shu et al.(1987)]{shu87}
 Shu, F. H., Adams, F. C., \& Lizano, S. 1987, \araa, 25, 23
\bibitem[Stepinski(1992)]{stepinski92}
 Stepinski, T. F. 1992, Icarus, 97, 130
\bibitem[Stone \& Norman(1992)]{stone92}
 Stone, J. M., \& Norman, M. L. 1992a, \apjs, 80, 753, 1992b, \apjs, 80, 791
 \bibitem[Toomre(1964)]{toomre64}
Toomre, A. 1964, \apj, 139, 1217
\bibitem[Umebayashi(1983)]{ume83}
 Umebayashi, T. 1983, Prog. Theor. Phys., 69, 480
\bibitem[Weidenschilling(1977)]{weiden77}
 Weidenschilling, S. J. 1977, \mnras, 180, 57
 \bibitem[Weidenschilling(1980)]{weiden80}
 Weidenschilling, S. J. 1980, Icarus, 44, 172
\bibitem[Wetherill \& Steward(1989)]{wetherill89}
 Wetherill, G. W., \& Steward, G. R., 1989, Icarus, 77, 330
\bibitem[Wisdom \& Tremaine(1988)]{wisdom88}
 Wisdom, J., \& Tremaine, S. 1988, \aj, 95, 925
\bibitem[Yabe \& Aoki(1991)]{yabe91}
 Yabe, T., \& Aoki, T. 1991, Comput. Phys. Comm., 66, 219
\bibitem[Youdin \& Chiang(2004)]{youdin_chinag04}
 Youdin, A. N. \& Chiang, E. I. 2004, \apj, 601, 1109
\bibitem[Youdin \& Shu(2002)]{youdin_shu02}
 Youdin, A. N. \& Shu, F. 2002, \apj, 580, 494
\bibitem[Youdin \& Goodman(2005)]{you05}
 Youdin, A. N., \& Goodman, J. 2005, \apj, 662, 613
\bibitem[Youdin \& Johansen(2007)]{you07}
 Youdin, A. N., \& Johansen, A. 2007, \apj, 662, 613
\bibitem[Zsom et al.(2010)]{zsom10}
 Zsom, A., Ormel, C. W., G\"{u}ttler, C., Blum, J., \& Dullemond, C. P. 2010, \aap, 513, A57

\end{thebibliography}
\end{document}